\documentclass[conference]{IEEEtran}
\IEEEoverridecommandlockouts

\usepackage{cite}
\usepackage{amsmath,amssymb,amsfonts}
\usepackage{algorithmic}
\usepackage{graphicx}
\usepackage{textcomp}
\usepackage{xcolor}
\usepackage{url}
\usepackage{graphicx}
\usepackage{xspace}
\usepackage{subcaption}
\def\BibTeX{{\rm B\kern-.05em{\sc i\kern-.025em b}\kern-.08em
    T\kern-.1667em\lower.7ex\hbox{E}\kern-.125emX}}
    
\begin{document}
\newcommand{\lm}[1]{\footnote{{\bf Luca: #1}}}
\newcommand{\tv}[1]{\footnote{{\bf Thiemo: #1}}}
\newcommand{\ws}[1]{\footnote{{\bf Weining: #1}}}
\newcommand{\sk}[1]{\footnote{{\bf Stefanos: #1}}}
\newcommand{\yy}[1]{\footnote{{\bf Yuan: #1}}}

\newcommand{\rv}{\textcolor{red}}
\newcommand{\rvt}{\textcolor{blue}}

\newcommand{\aem}[0]{\textsc{AEM}\xspace}
\newcommand{\ar}[0]{\textsc{AR}\xspace}
\newcommand{\cem}[0]{\textsc{CEM}\xspace}

\newcommand{\co}[0]{\textsc{Check-Only}\xspace}

\newcommand{\capt}[1]{\mdseries{\emph{#1}}}
\newcommand{\code}[1]{\textbf{{\texttt{#1}}}}
\newcommand{\fakepar}[1]{\vspace{.5mm}\noindent\textbf{#1.}}
\newcommand\figref[1]{Fig.\,\ref{#1}}
\newcommand\secref[1]{Sec.\,\ref{#1}}
\newcommand\tabref[1]{Tab.\,\ref{#1}}
\newcommand\listref[1]{List\,\ref{#1}}
\title{Mobile Backscatter Communication\\ for the Battery-less Internet of Things}

\makeatletter
\newcommand{\linebreakand}{%
  \end{@IEEEauthorhalign}
  \hfill\mbox{}\par
  \mbox{}\hfill\begin{@IEEEauthorhalign}}
\makeatother

\author{\IEEEauthorblockN{Weining Song}
\IEEEauthorblockA{
\textit{Uppsala University}\\
Sweden \\
weining.song@angstrom.uu.se}
\and
\IEEEauthorblockN{Thiemo Voigt}
\IEEEauthorblockA{
\textit{Uppsala University, RISE}\\
Sweden \\
thiemo.voigt@angstrom.uu.se}
\and
\IEEEauthorblockN{Stefanos Kaxiras}
\IEEEauthorblockA{
\textit{Uppsala University}\\
Sweden \\
stefanos.kaxiras@it.uu.se}
\linebreakand
\IEEEauthorblockN{Yuan Yao}
\IEEEauthorblockA{
\textit{Uppsala University}\\
Sweden \\
yuan.yao@it.uu.se}
\and
\IEEEauthorblockN{Luca Mottola}
\IEEEauthorblockA{
\textit{Politecnico di Milano, RISE, Uppsala
University}\\
Italy and Sweden \\
luca.mottola@angstrom.uu.se}

}

\maketitle

\begin{abstract}

We enable backscatter communication in the \emph{battery-less mobile} Internet of Things (IoT).
Backscatter communication is extensively studied in static settings. 
Existing designs are, however, fundamentally mismatched with mobility and time-varying energy patterns.
Channel conditions rapidly fluctuate, impacting the achievable data rates and thus transmission costs. 
Energy availability varies unpredictably, possibly forcing devices to remain quiescent to recharge energy buffers.
The two issues compound each other: while recharging, a battery-less mobile IoT device may miss more favorable channel conditions.
We design a lightweight decision system that dynamically determines when to transmit by checking short-term trends in signal strength, while using Non-volatile Memory (NVM) to retain packets in unfavorable channel conditions and across energy failures.
Using a prototype we built and real-world mobility and power traces, we compare our design against a rate-adaptive baseline that only considers the instantaneous channel conditions.
Experimental results show that our system improves throughput by up to 5.16\,$\times$ while reducing transmission energy consumption by up to 47.3\%, with only 0.23\% -- 7.3\% additional energy overhead.

\end{abstract}


\section{Introduction}
Backscatter communication allows battery-less Internet of Things (IoT) devices to transmit data with minimal energy consumption by modulating information onto an ambient carrier signal~\cite{liu2013ambient,van2018ambient}. 
Most existing backscatter systems are designed under the assumption of static environments, where channel conditions remain stable~\cite{kellogg2014wi,peng2018plora,wang2017fm}.
However, many IoT applications involve mobile energy-harvesting devices operating in dynamic environments~\cite{ahmed2024internet,khalifa2017harke,wang2022overview,lai2022self,haroun2021progress}, where both channel conditions and harvested energy vary rapidly over time.
The mismatch between static design assumptions and dynamic operating conditions fundamentally challenges the performance of existing backscatter systems.

\fakepar{Mobility in backscatter communications}
Device mobility leads to continuously varying channel conditions between the backscatter tag, the carrier emitter, and the receiver~\cite{muratkar2021ambient}.
For example, channel conditions degrade with increasing distance from the carrier emitter, requiring the system to transmit at a lower data rate to retain packet reliability~\cite{huang2021freeback}.

\begin{figure} [tb]
\centering
\includegraphics[width=0.88\linewidth]{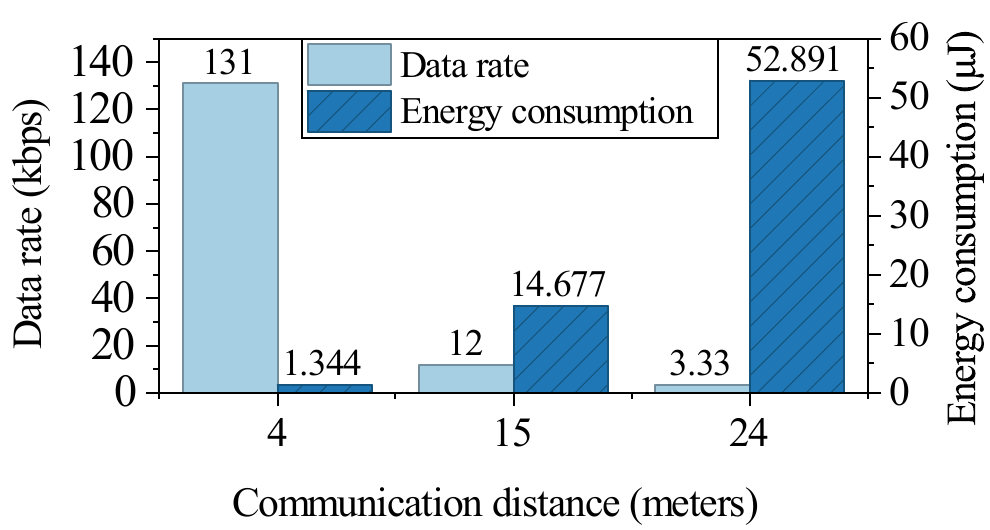}
\caption{Achievable data rate and per-packet energy consumption versus communication distance.
\capt{Transmission energy increases with reduced data rate at longer distances.}} 
\label{fig:Mov_intro}
\vspace{-5mm}
\end{figure}

\figref{fig:Mov_intro} provides quantitative evidence from our own experimental measurements, obtained by replicating the setup of prior work~\cite{huang2021freeback}. As the tag moves away from the co-located carrier emitter and receiver, the achievable data rate decreases accordingly.
Consequently, each packet requires a longer transmission time, leading to higher energy consumption.
\figref{fig:Mov_intro} shows that a data rate of 131\,kbps is attainable at a 4-meter communication distance from the co-located emitter and receiver, with an energy cost of only 1.344\,$\mu$J per packet.
However, to transmit the same data over a 24-meter communication distance, the energy requirement surges by 39 times due to a reduction in data rate, prolonged transmission times, and extended microcontroller active time.

At the same time, energy harvested from the environment is erratic and scarce, possibly causing frequent energy failures that force the device to stay off while recharging energy buffers. 
This generally results in a time-varying energy budget for transmission, which is difficult to predict.
Worse, when the device is off, recharging and yet still mobile, possibly favorable channel conditions are missed~\cite{ahmed2024internet,bhatti2016energy}.

When mobility and energy harvesting coexist, both channel conditions and available energy vary over time and influence each other, leading to a distinct design problem where transmission decisions depend on future channel and energy dynamics.
This limits the effectiveness of pre-tuned transmission strategies and motivates the need for runtime adaptation.

\fakepar{Mobile transmission control} Under time-varying channel conditions and energy availability, the system should intelligently decide \emph{when it is the right time to transmit}. 
For example, provided sufficient energy is available, the system may decide to postpone a transmission in case channel conditions are poor, trading packet latency for the reduced energy consumption that transmitting at a later time, with better channel conditions and thus higher data rates, would allow.
Doing so, given the stochastic nature of the time-varying quantities at hand is difficult.
Solutions to the problem should, moreover, co-exist with the scarce compute resources that we expect mobile backscatter IoT devices to be equipped with, and thus impose minimal overhead.

Unlike existing designs that base transmission decisions on instantaneous signal strength~\cite{huang2021freeback}, our system monitors how the received signal strength indicator (RSSI) at the backscatter tag changes over time. 
Depending on whether the signal strength is increasing or decreasing, the system adjusts transmission decisions accordingly. 
When channel conditions improve and sufficient energy is available, the system provides more opportunities for packet transmissions. 
When conditions worsen, packet transmissions are throttled down to conserve energy for future better opportunities.
We use modern Non-Volatile Memory (NVM) as a temporary storage facility to persist unsent packets across energy failures and periods when the channel conditions are poor, allowing transmission to resume once energy and channel conditions improve.
This design enables the device to adapt to time-varying channel and energy conditions, ultimately improving overall throughput.

We prototype the system on an MSP430FR5969 MCU and implement three representative IoT applications: agricultural environment measurement (\aem), activity recognition (\ar), and cold-chain equipment monitoring (\cem). 
These applications differ in processing workload and packet size, leading to different demands on computation, buffering, and transmission.
We use a trace-driven simulator to evaluate the system and compare it with a rate-adaptive baseline~\cite{huang2021freeback}.
Using real-world signal traces collected in our environment and power traces from prior work~\cite{geissdoerfer2022learning}, experimental results show that our approach achieves up to 5.16\,$\times$ the long-term throughput of the baseline. 
This improvement is enabled by up to 47.3\% reduction in transmission energy consumption compared to the baseline, while introducing only 0.23\% -- 7.3\% additional energy overhead.

The rest of the paper is organized as follows.
\secref{sec:background} provides background information and discusses related work.
\secref{sec:sys} presents the system design and implementation.
\secref{sec:eva} presents the hardware prototype, benchmarks, baselines, evaluation setup, and experimental results.
\secref{sec:conclusion} ends the paper.

\section{Background and Related Work}
\label{sec:background}
We provide background information useful for the remainder of the paper and briefly survey related work next.

\fakepar{Intermittent computing}
Battery-less IoT devices use ambient energy sources as their only power source~\cite{bhatti2016energy}. 
This leads to frequent energy failures during program execution due to erratic energy patterns, resulting in an intermittent computing pattern~\cite{ahmed2024internet}.
Consequently, active operation is often interrupted by periods of recharging energy buffers.

To maintain forward progress, these devices must leverage NVM to persist program state across energy failures.
This state is restored when the energy buffer fills up, allowing execution to resume close to the energy failure point rather than performing a complete reboot.
Some solutions use checkpointing to save the intermediate program state to NVM~\cite{ahmed2019efficient,balsamo2014hibernus,balsamo2016hibernus++,bhatti2017harvos,maeng2018adaptive,ransford2011mementos,van2016intermittent}. 
This approach replicates the current state at specific points in the program and restores it once the system has enough energy. 
Existing techniques differ in how they place these checkpoints and whether they decide the checkpoint locations at compile-time or run-time.

These techniques primarily focus on ensuring correct program execution across power failures. 
They do not consider using NVM to proactively regulate communication decisions.
In contrast, our work uses NVM to persist unsent packets and control state, allowing the system to proactively store packets even when sufficient energy is available and transmit them later under more favorable channel conditions.

\fakepar{Backscatter communication}
By selectively reflecting or absorbing radio frequency (RF) signals rather than generating its own carrier, backscatter communications enable data transmission with $\mu\mathrm{W}$ power consumption~\cite{varshney2017lorea,liu2013ambient,kellogg2014wi,wang2017fm,talla2017lora,bharadia2015backfi,peng2018plora}.

Close to our work is the design of Xu et al., who present a battery-free backscatter system that relies on intermittent WiFi signals for energy harvesting and communication~\cite{xu2020ecuwb}.
This work predicts excitation duration and schedules charging and transmission accordingly.
In contrast, we consider scenarios where device mobility causes continuously varying channel conditions and fluctuating harvested energy, rather than relying on a stable carrier with unpredictable duration.

Muratkar et al. study ambient backscatter communication with a mobile RF source and analyze the impact of mobility on physical-layer performance through channel modeling and detection analysis~\cite{muratkar2021ambient}.
In contrast, our work focuses on system-level transmission control under mobility and energy harvesting, rather than physical-layer characterization.

FreeBack~\cite{huang2021freeback} is a backscatter communication system that adapts the transmission rate according to channel conditions.
The tag estimates channel quality from the incoming carrier signal and adjusts the data rate accordingly.
Unlike our work, FreeBack makes transmission decisions based solely on instantaneous signal strength without considering energy availability or proactively regulating transmissions.

\section{System Design}
\label{sec:sys}

We articulate the system design and elaborate on its features, including transmission control and energy management.

\subsection{Intuition}
\label{sec:mov}

Most existing work on backscatter communication assumes a static environment with relatively stable channel conditions~\cite{kellogg2014wi,peng2018plora,wang2017fm}.
However, in many IoT applications, the backscatter tag is mobile and powered by energy harvesting, resulting in continuous fluctuations in both channel conditions and energy availability~\cite{ahmed2024internet,khalifa2017harke,wang2022overview,lai2022self,haroun2021progress}.

Huang et al.'s backscatter system~\cite{huang2021freeback} is an example design conceived for handling channel variations.
It adjusts the transmission rate according to the instantaneous signal strength.
In this work, the carrier emitter and receiver are co-located, which simplifies channel estimation based on the received signal strength of the carrier at the tag.

We replicate the experimental setup~\cite{huang2021freeback} to characterize the relationship between RSSI, communication distance, data rate, and energy consumption.
The results in \figref{fig:Freeback} show that when using the MSP430FR5969 MCU to generate a 256-bit packet, the data rate drops from 1232\,kbps at -14\,dBm to 0.33\,kbps at -46\,dBm, while energy consumption increases from 0.14\,$\mu$J to 533.72\,$\mu$J, which is over 3000\,$\times$ higher.
Thus, transmitting under poor channel conditions can waste scarce harvested energy.
Furthermore, the rate-adaptive system only considers instantaneous channel conditions, ignoring both cases where the tag moves out of communication range and where energy failures make transmission impossible in mobility and energy-harvesting scenarios.

\begin{figure}[tb]
\centering
\includegraphics[width=0.85\linewidth]{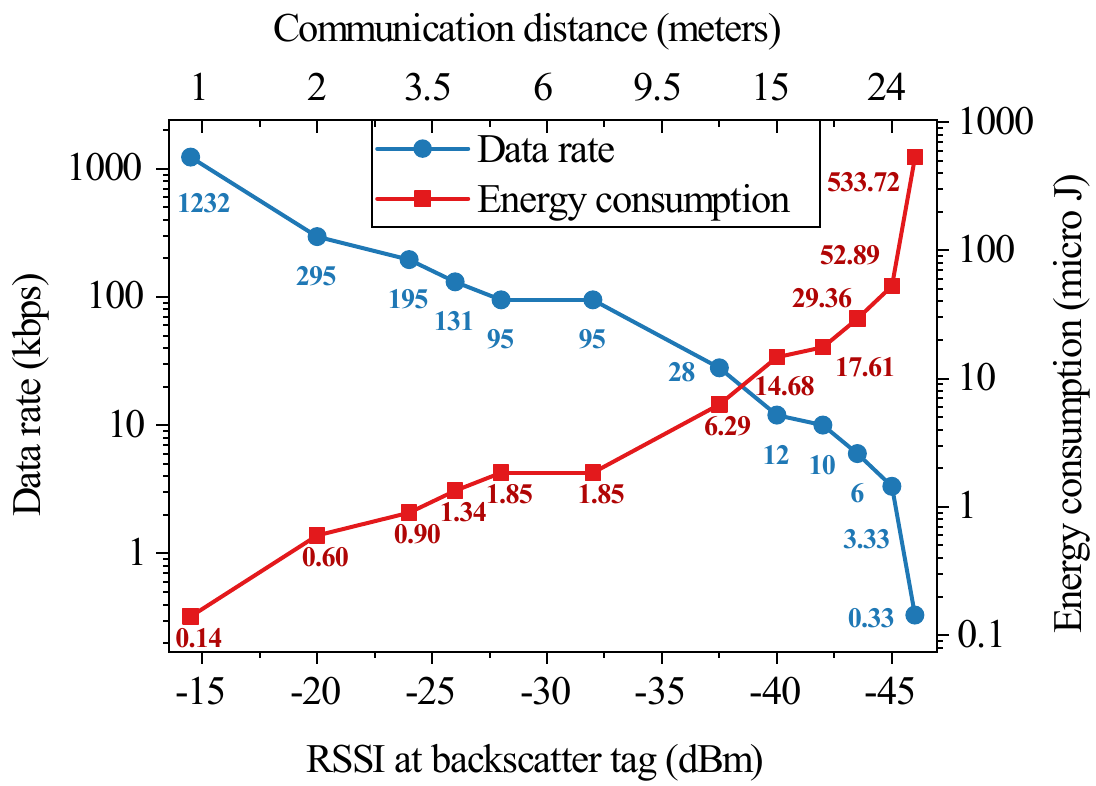}
\caption{Impact of channel quality and communication distance on transmission rate and energy cost~\cite{huang2021freeback}.}
\label{fig:Freeback}
\vspace{-6mm}
\end{figure}

In such scenarios, both future energy input and channel conditions are unknown, leading to a fundamental issue: \emph{when it is the right time to transmit}.
\figref{fig:Mov_a} illustrates the issue.
At 2.5\,s, the capacitor is fully charged, and the system transmits at an RSSI of -28 \,dBm.
A single packet transmission consumes significant energy, causing the capacitor voltage to drop from 4.0\,V to 2.2\,V.
This results in a long recharging period from 2.5\,s to 12\,s, missing the strong signal window (7\,s -- 9\,s).

Our design is exemplified in~\figref{fig:Mov_t}. 
The application still generates a packet at 2.5\,s, but defers its transmission and stores it in NVM, where packets generated by the application in unfavorable channel conditions are buffered over time.
As a result, the tag preserves its energy and maintains a higher voltage level, allowing it to resume transmission sooner as the signal strength improves.
When the capacitor is fully charged again, the corresponding RSSI is -19\,dBm.
Under the same energy budget, where the capacitor voltage drops from 4.0\,V to 2.2\,V, our system transmits three packets in a burst: two buffered in NVM from earlier weak channel conditions, and one newly generated packet.
By aligning transmission with favorable channel conditions, we achieve significantly higher throughput with the same energy budget.

\begin{figure*}[tb]
    \centering
    \begin{subfigure}{0.45\linewidth}
        \centering
        \includegraphics[height=5.8cm]{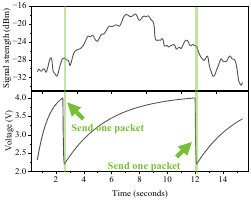}
        \caption{Rate-adaptive backscatter system.}
        \label{fig:Mov_a}
       \vspace{-2mm}
    \end{subfigure}
    \hfill 
    \begin{subfigure}{0.45\linewidth}
        \centering
        \includegraphics[height=5.8cm]{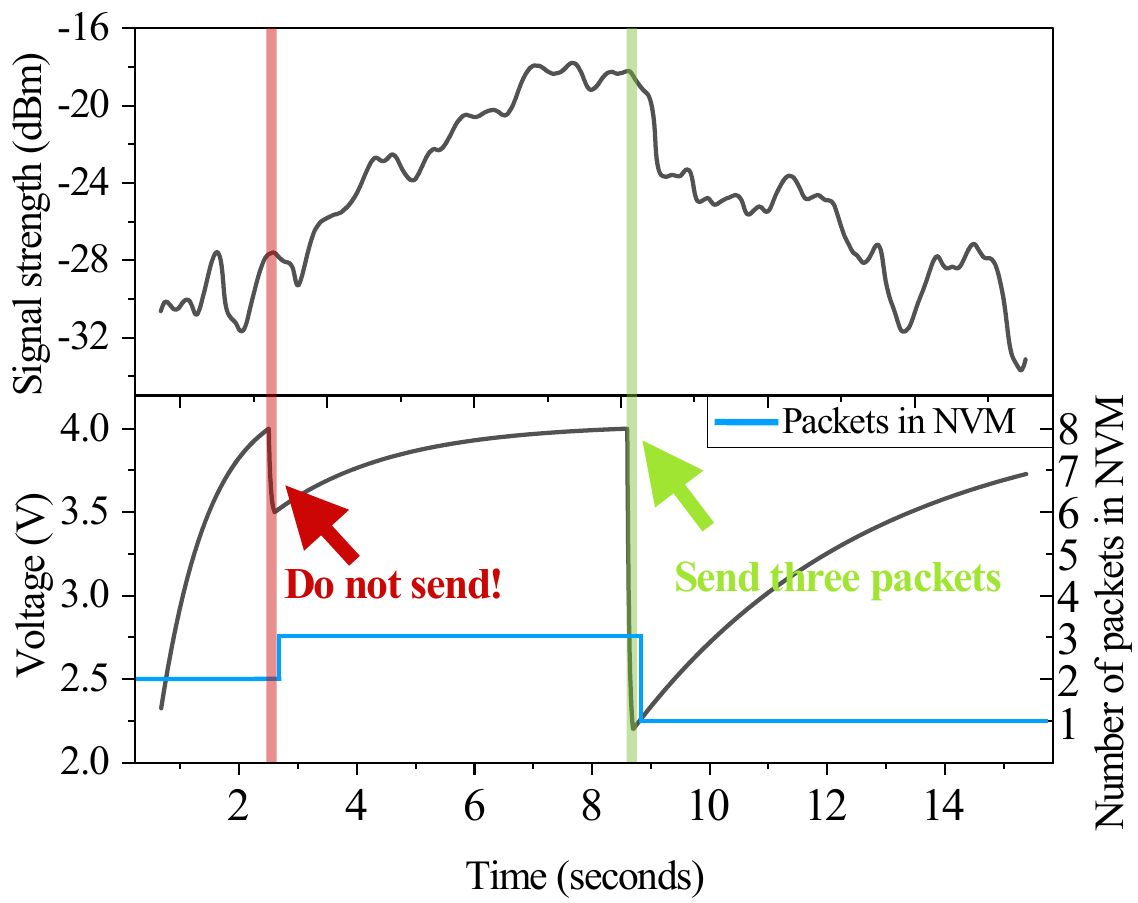}
        \caption{Mobile backscatter system.}
        \label{fig:Mov_t}
      \vspace{-2mm}
    \end{subfigure}
    \caption{Example comparing rate-adaptive and mobile backscatter systems under identical signal and power traces.
    \capt{Mobile backscatter system achieves higher transmission efficiency under the same conditions.}}
    \label{fig:Mov}
    \vspace{-3mm}
\end{figure*}

Ultimately, the key challenge is to decide when and how aggressively the system should transmit to maximize throughput under time-varying channel and energy conditions.

\begin{figure}[tb]
\centering
\includegraphics[width=0.75\linewidth]{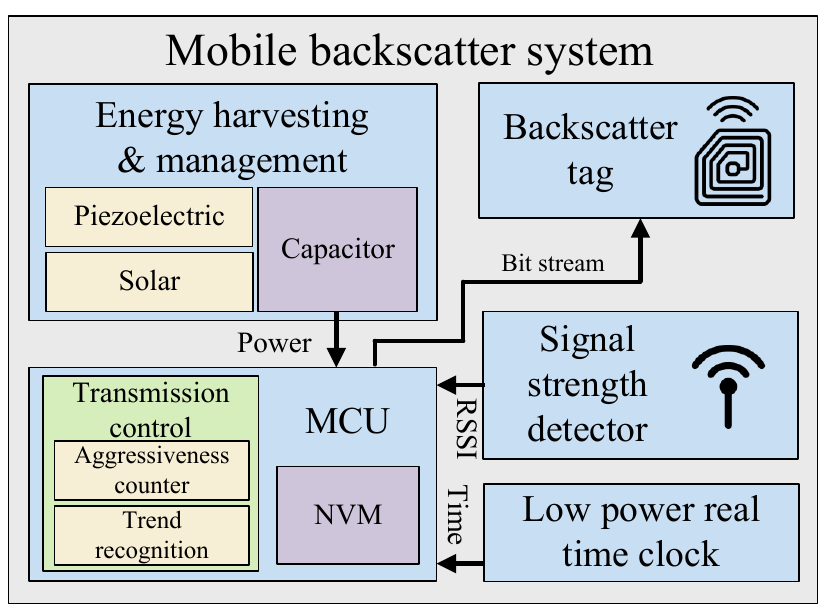}
\caption{Mobile backscatter communication system.}
\label{fig:system}
\vspace{-6mm}
\end{figure}

\subsection{System Overview}

Our design is loosely inspired by TCP (Transmission Control Protocol) congestion control, where each source estimates the available network capacity to determine how many packets can be safely transmitted~\cite {allman2009tcp}. 
Rather than estimating network capacity, our system infers future transmission opportunities from channel trends and energy conditions, and adjusts transmission schedules accordingly.

\figref{fig:system} provides an overview of the architecture of our system.
It consists of an MCU with on-chip NVM, a signal strength detector, a backscatter tag, a low-power real-time clock (RTC), and an energy harvesting and management circuit with a capacitor for temporary energy storage.
We implement a lightweight, runtime transmission control mechanism on the MCU.
By monitoring the RSSI and estimating its trend after each energy recharge cycle, the transmission controller decides whether to store packets in the NVM or initiate transmission.

\subsection{Transmission Control}

\fakepar{RSSI trend recognition}
To detect channel trends with minimal state and computational overhead, the system maintains an exponential moving average (EMA) of the RSSI signal at the tag as a smoothed reference of recent channel conditions.
The EMA is updated as follows:
\begin{equation}
S_n = \alpha_n R_n + (1 - \alpha_n) S_{n-1},
\end{equation}
where $R_n$ is the current RSSI sample, $S_n$ is the smoothed estimate, and $S_{n-1}$ is the previous estimate.
Since the system operates under energy harvesting, device wake-ups and RSSI sampling occur at irregular intervals: $\alpha_n$ is a smoothing factor, which is adapted based on the time interval between consecutive samples, $\Delta t$, measured by the local RTC:
\begin{equation}
\alpha_n = 1 - e^{-\Delta t / \tau},
\end{equation}
where $\tau$ controls the effective memory of the filter, determining how much past RSSI samples influence the reference estimate, and is set to 5 s in all experiments.
This allows the EMA to naturally adjust to irregular sampling while still providing a stable reference.

The instantaneous RSSI sample $R_n$ reflects the channel quality from the emitter to the tag.
In our setup, the carrier emitter and receiver are co-located, allowing this measurement to be used to estimate the overall backscatter channel quality.
We obtain the current channel trend by subtracting the EMA reference, $S_n$, from the instantaneous RSSI sample:
\begin{equation}
T_n = R_n - S_n,
\end{equation}
where a positive $T_n$ indicates increasing channel quality, while a negative value indicates decreasing channel quality.
This compares the current RSSI against the EMA reference to capture short-term channel dynamics.

This approach is computationally inexpensive and requires a small memory footprint.  
It provides sufficient trend information to guide transmission decisions without relying on complex prediction models for future signal estimation.

\fakepar{Transmission budget}
The system maintains a non-negative aggressiveness counter that determines the number of packets scheduled for transmission during each active period.
The counter is initialized to zero and updated at the beginning of each active period based on the detected RSSI trend and the observed charging duration from the RTC.

If the RSSI trend is increasing, the system increases the aggressiveness counter to take advantage of improving channel conditions.
The increment step is the most recent charging duration. 
A longer charging duration implies low input energy, leading to less frequent active phases.
Consequently, each active period should transmit more packets to compensate for the reduced availability of future active periods due to limited harvested energy.
Conversely, a shorter duration implies higher input energy and more frequent transmission opportunities in the future.
In this case, only small counter adjustments are required, as the system can distribute packet transmissions across multiple upcoming opportunities instead of concentrating them within a single active phase.

Instead, if the RSSI trend is decreasing, the counter is reduced using the same charging-time-dependent step size. 
Once the counter reaches zero, the system proactively stores packets in the NVM, which is managed as a FIFO buffer, allowing them to be sent later when channel conditions improve.
This ensures that under decreasing channel conditions, the system reduces the aggressiveness counter while considering fewer future transmission opportunities.

By integrating channel trends with the observed recent charging history, the system dynamically adjusts its transmission aggressiveness. 
This update rule coordinates packet transmissions with time-varying channel conditions and energy availability, enabling transmissions to exploit favorable conditions while avoiding unnecessary energy consumption.

\fakepar{Energy management}
The system executes on unstable harvested energy, which is buffered in a capacitor.
When the capacitor voltage drops below a prespecified threshold, the MCU shuts down, interrupting ongoing tasks. 

To ensure forward progress, the system persists critical state and unsent packets in NVM when transmission is not possible due to low energy or poor channel conditions. 
In addition, it stores a minimal transmission control state, which includes the state of the aggressiveness counter and the RSSI trend recognition algorithm.
When energy is restored, the MCU resumes execution by recovering the persisted packets and control state from NVM.
This mechanism ensures intermittent computing without losing data, despite frequent energy failures and time-varying channel conditions.

\section{Evaluation}
\label{sec:eva}

In this section, we present the prototype used for evaluation, benchmarks, baselines, and the experimental setup, followed by an evaluation of system throughput under real-world signal and power traces.
We then analyze energy consumption and the overhead of transmission control. 
Finally, we use two microbenchmarks to evaluate how different system factors and parameter settings affect throughput.

Our results show our design achieves up to 5.16\,$\times$ the throughput of a rate-adaptive baseline, by reducing transmission energy consumption by up to 47.3\%, while introducing only 0.23\% -- 7.3\% additional energy overhead.

\subsection{Prototype}
\label{Prototype}
We prototype the mobile backscatter communication system using off-the-shelf hardware. 
The controller is implemented on an MSP430FR5969 MCU~\cite{MSP430FR5969}, a low-power microcontroller with 64\,KB of on-chip FRAM. 
The MCU executes application logic on top of the transmission control layer of \secref{sec:sys}.

For signal strength sensing, we employ a TI CC1352P7~\cite{CC1352P7} to obtain RSSI measurements, which are forwarded to the MSP430FR5969 via the serial peripheral interface (SPI). 
The backscatter tag is built upon the LoRea architecture~\cite{varshney2017lorea}. 
An EM3028-C7 RTC is used to track time and support time-adaptive trend recognition and aggressiveness control~\cite{EM3028-C7}.
It also provides timestamping for application-layer data.

\subsection{Benchmarks}
\label{sec:applications}
We deploy three representative IoT applications to evaluate the performance of the mobile backscatter system.

\fakepar{Agricultural environment measurement (\aem)}
\aem measures the temperature difference between ambient air and plant leaf surfaces using high-precision sensors~\cite{daskalakis2017ambient}.
The temperature difference and timing information are periodically transmitted as a 256-bit packet, representing a computation-intensive sensing workload with periodic reporting.

\fakepar{Activity recognition (\ar)}
\ar classifies device motion based on a three-axis accelerometer using a lightweight nearest-neighbor classifier~\cite{maeng2017alpaca}.
Only the inference result and essential metadata, including timestamp and sequence information, are transmitted, resulting in a 128-bit packet. 
This represents a lightweight computation and communication demand.

\fakepar{Cold-chain equipment monitoring (\cem)}
\cem logs temperature data for cold-chain storage and transport~\cite{maeng2017alpaca}. 
Given the relatively stable temperature patterns, \cem applies LZW compression before transmission. 
Instead of reporting each new reading immediately, data are buffered and compressed in blocks using a 512-entry dictionary and a 64-byte compressed block size, and a block is transmitted once the buffer is full as a 640-bit packet. 
This reflects a monitoring application with compression cost and bursty transmission behavior.
\begin{figure}[tb]
\centering
\includegraphics[width=0.9\linewidth]{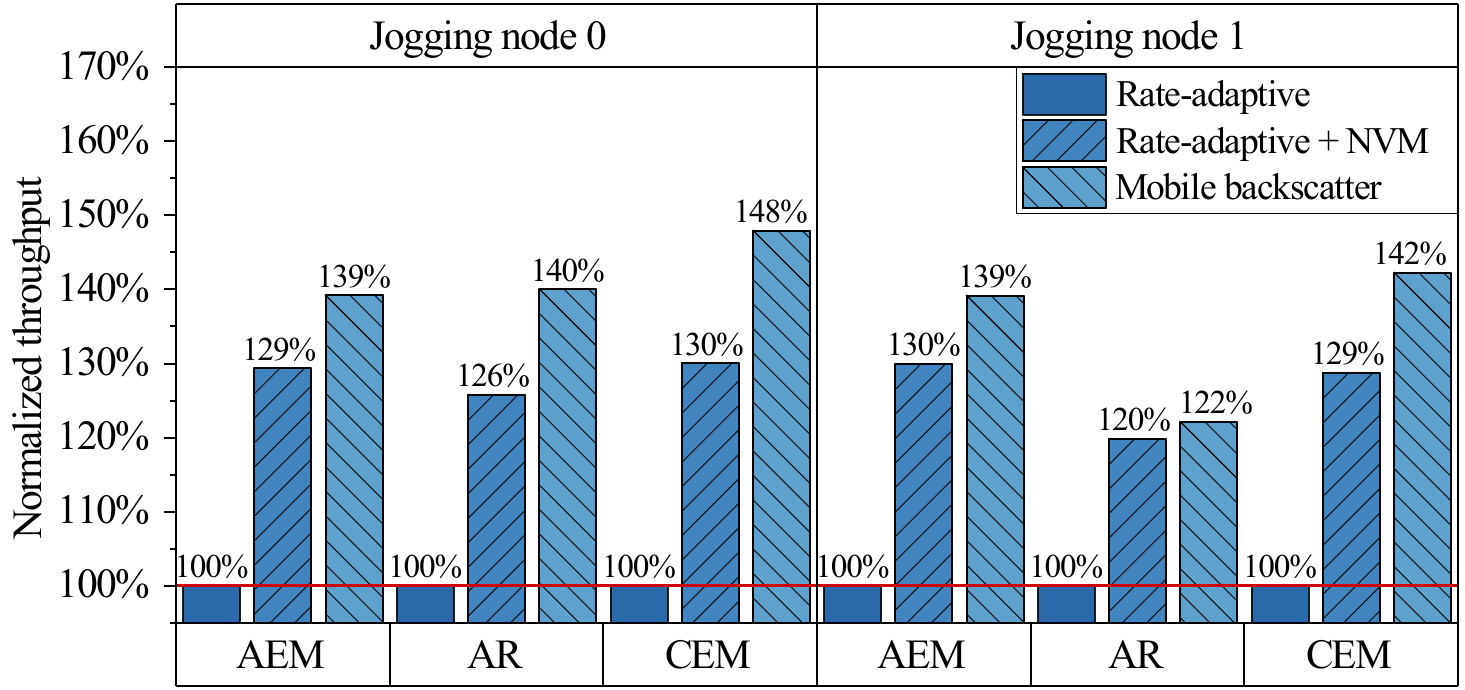}
\caption{Normalized average throughput. \capt{The mobile backscatter system achieves up to 1.48\,$\times$ average throughput.}}
\label{fig:Overall_improve}
\vspace{-6mm}
\end{figure} 

\begin{figure*}[tb]
\centering
\includegraphics[width=1\linewidth]{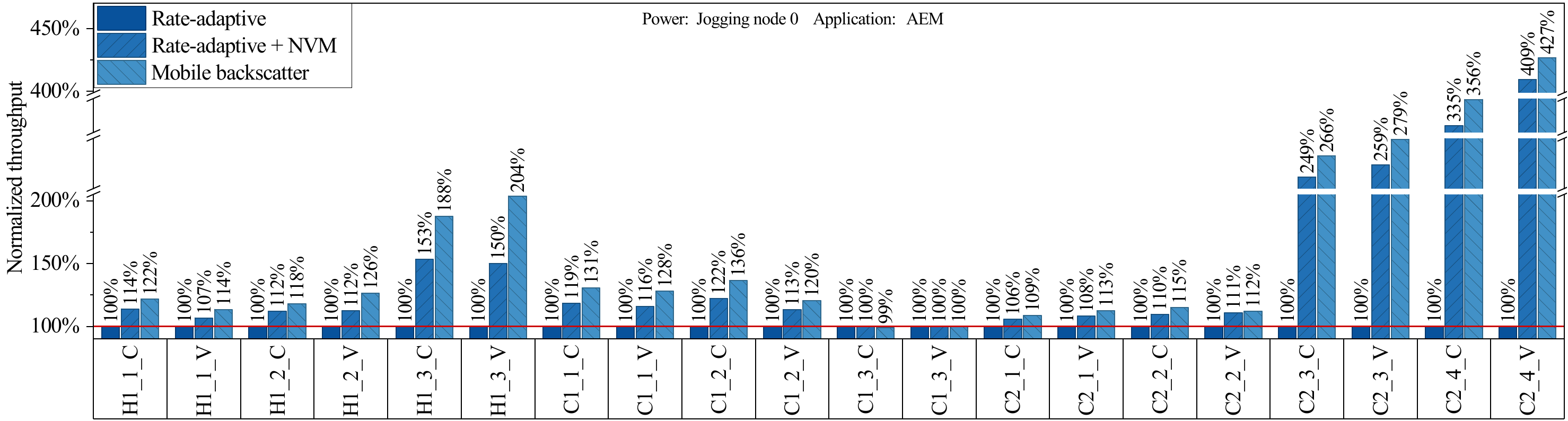}
\caption{Normalized throughput per signal trace for the \aem application under the jogging node 0  power trace. \capt{The largest improvements occur when the RSSI is most of the time below the transmission threshold and transmissions can be deferred until conditions improve (e.g. \textit{C2\_4\_V}).}}
\label{fig:Improve_AEM}
\vspace{-3mm}
\end{figure*}

\begin{figure*}[tb]
\centering
\includegraphics[width=1\linewidth]{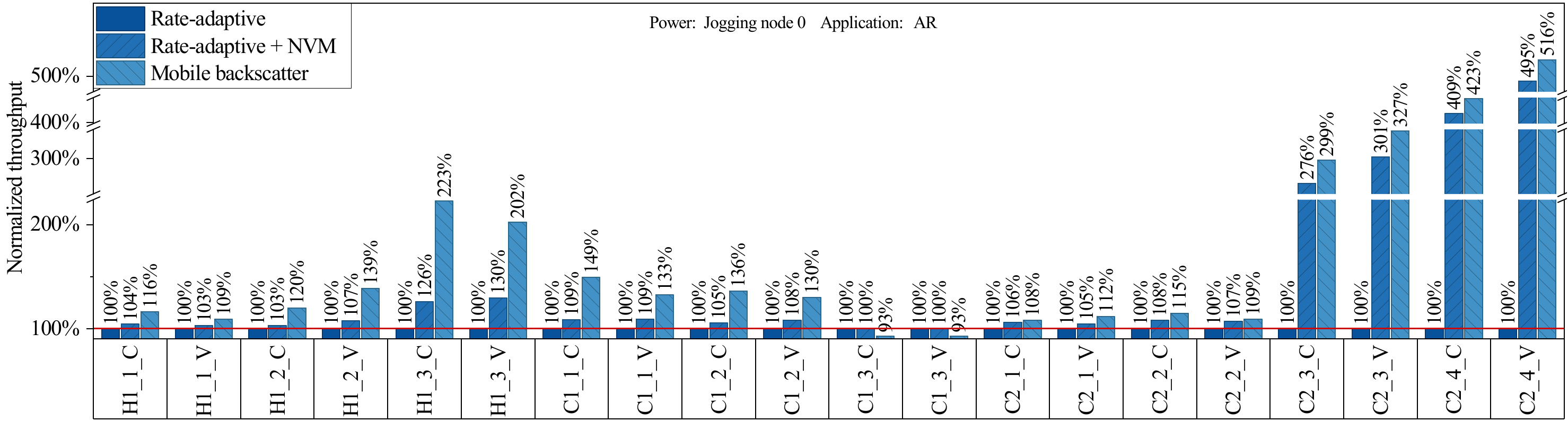}
\caption{Normalized throughput per signal trace for the \ar application under the jogging node 0 power trace.}
\label{fig:Improve_AR}
\vspace{-3mm}
\end{figure*}

\begin{figure}[tb]
\centering
\includegraphics[width=0.9\linewidth]{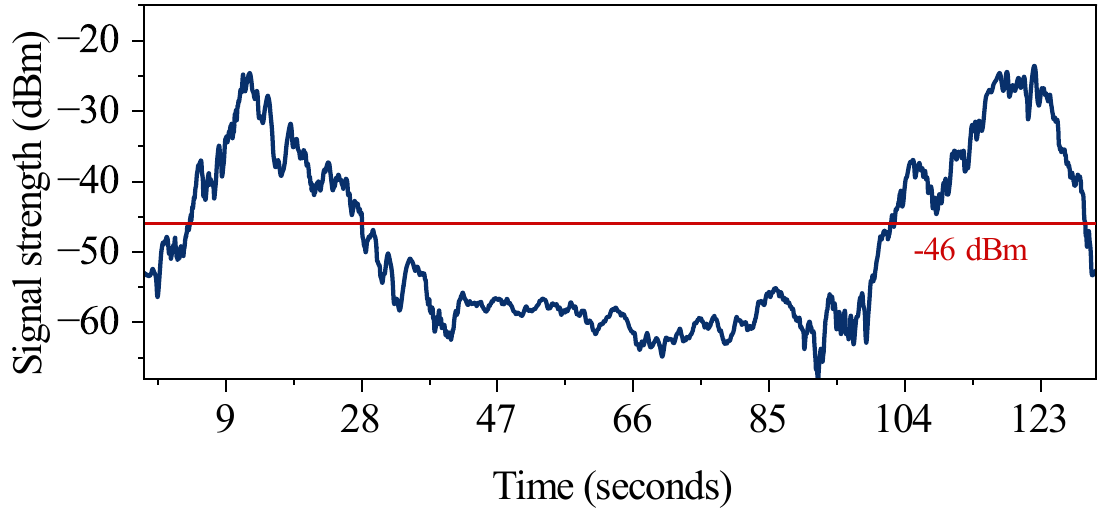}
\caption{The RSSI recorded at corridor 2, walking path 4, under varying speed (C2\_4\_V).
\capt{The RSSI falls below the transmission threshold for 63.4\% of the time.}}
\label{fig:C2_4_V}
\vspace{-4mm}
\end{figure}

\begin{figure*}[tb]
\centering
\includegraphics[width=1\linewidth]{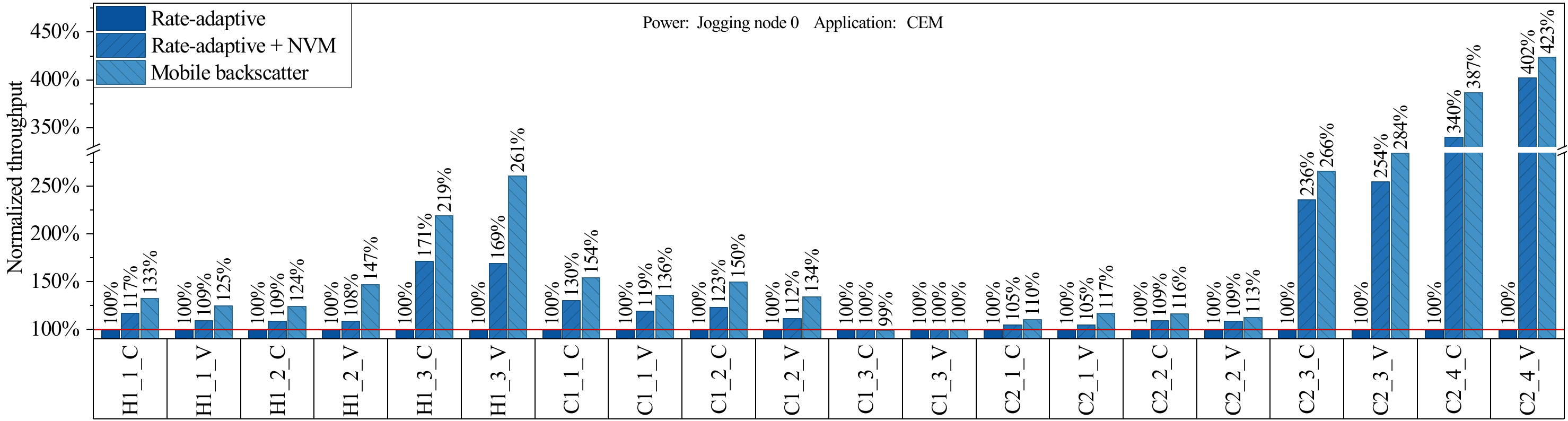}
\caption{Normalized throughput per signal trace for the \cem application under the jogging node 0 power trace.}
\label{fig:Improve_CEM}
\vspace{-3mm}
\end{figure*} 

\begin{figure*}[tb]
\centering
\includegraphics[width=1\linewidth]{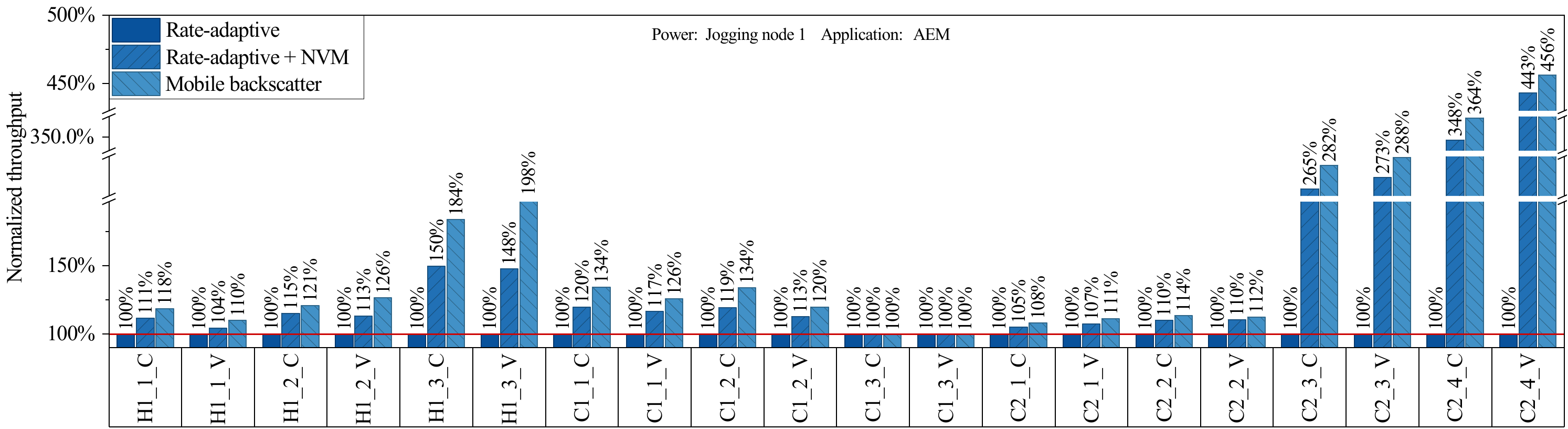}
\caption{Normalized throughput per signal trace for the \aem application under the jogging node 1 power trace.}
\label{fig:Improve_AEM_node1}
\vspace{-3mm}
\end{figure*} 

\subsection{Experimental Setup}
In our experiments, the carrier emitter and receiver are co-located and remain stationary, while the backscatter tag is mobile, reflecting realistic IoT scenarios with devices moving relative to a fixed infrastructure.
The tag uses a capacitor sized according to the energy demand of the running application, ensuring sufficient energy to complete at least one active period.

Since running the prototype for extended periods under reproducible signal and energy environments is impractical, we build a measurement-driven simulation framework to evaluate long-term system behavior under precisely controlled scenarios.
To build an accurate model, we run applications on the hardware prototype in~\secref{Prototype} and measure energy consumption and execution time. 
The prototype is connected to a digital power supply for power measurement and to an oscilloscope for execution timing. 
These measurements provide precise task-level energy and timing profiles.
Based on these measurements, the simulator models capacitor charging and discharging, task execution timing, backscatter transmission behavior, and the corresponding energy consumption. 

The simulator takes the recorded power and signal traces as inputs, replaying energy harvesting and RSSI variations caused by device mobility.
We use two real-world power traces collected from people jogging, representing realistic mobile energy harvesting conditions~\cite{geissdoerfer2022learning}. 
Multiple solar panels and piezoelectric harvesters are attached to two participants, and each trace has a duration of one hour.
We also record 20 real-world signal traces in diverse indoor environments, including two corridors and a large hall. 
These traces are collected from 10 movement paths with both constant-speed and varying-speed mobility.
Each trace lasts around two minutes and is extended to one hour by repeating it. 
We name each trace based on its location, path index, and speed.
For example, \textit{C1\_2\_C} denotes corridor 1, path 2, with constant speed.

\subsection{Baselines}
We compare our system against two baselines lacking transmission control to evaluate its effectiveness.

\fakepar{Rate-adaptive}
A rate-adaptive transmission system, based on the design of Huang et al.~\cite{huang2021freeback}, dynamically adjusts the data rate according to the instantaneous signal strength, as shown in~\figref{fig:Freeback}.
However, no mechanisms are employed to preserve packets when the RSSI falls below the transmission threshold of -46\,dBm in our prototype, or during energy failures, making it sensitive to mobility and energy harvesting scenarios.

\fakepar{Rate-adaptive + NVM}
The NVM-based rate-adaptive baseline extends the rate-adaptive baseline by using NVM to store packets that cannot be transmitted due to weak signals or energy failures. 
While this reduces packet loss, it does not leverage channel trends or energy status for transmission control. 
Therefore, energy may be spent inefficiently, limiting throughput under varying channel and energy conditions.

\subsection{Throughput}
\label{sec:throughput}
To understand how our transmission control improves throughput under mobility and energy harvesting environments, we measure the number of complete transmitted packets over a one-hour simulation for each application.

\fakepar{Average throughput}
\figref{fig:Overall_improve} shows the average throughput normalized to the rate-adaptive baseline.
On average, the mobile backscatter system can achieve 1.39\,$\times$ the throughput of the rate-adaptive baseline for the \aem application under both power traces.
For the \ar application, it achieves 1.40$\times$ and 1.22$\times$ the throughput over the rate-adaptive baseline under jogging node 0 and jogging node 1 power traces, respectively.
The \cem application achieves the highest throughput gains, reaching 1.48$\times$ and 1.42$\times$ the throughput over the baseline under the two different power traces.
We provide a detailed analysis of the factors affecting performance in the following section.
Overall, the results demonstrate the robustness of the mobile backscatter system across diverse applications with varying computational and communication demands, as well as under different signal and power traces.

\begin{figure}[tb]
\centering
\includegraphics[width=0.9\linewidth]{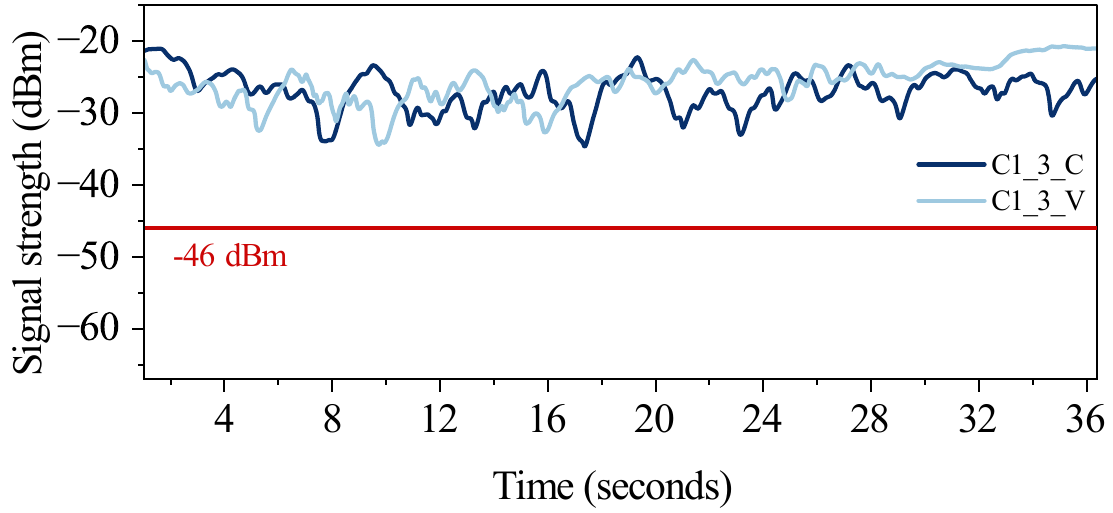}
\caption{The RSSI recorded at corridor 1, path 3, under constant and varying speeds (C1\_3\_C and C1\_3\_V).
\capt{RSSI fluctuates between -20\,dBm and -30\,dBm without a clear trend.}}
\label{fig:C1_3}
\vspace{-3mm}
\end{figure} 

\begin{figure*}[tb]
\centering
\includegraphics[width=1\linewidth]{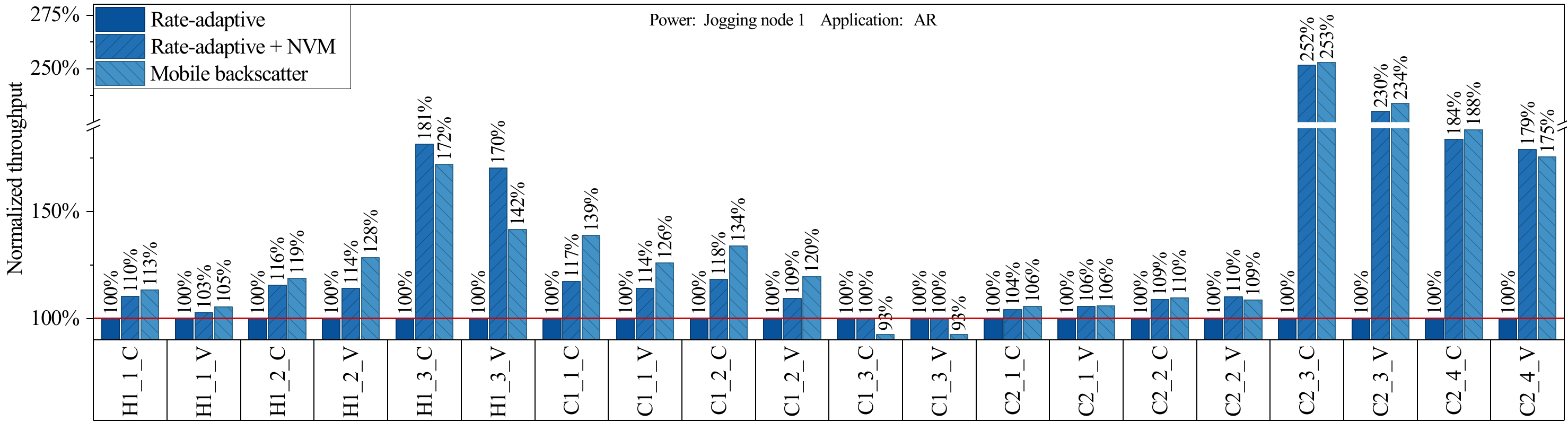}
\caption{Normalized throughput per signal trace for the \ar application under the jogging node 1 power trace.
\capt{In general, the mobile backscatter system improves the throughput. However, it achieves lower throughput than the NVM-based rate-adaptive baseline under some signal traces, due to insufficient transmission opportunities and excessive packet buffering in NVM.}}
\label{fig:Improve_AR_node1}
\vspace{-3mm}
\end{figure*} 

\begin{figure*}[tb]
\centering
\includegraphics[width=1\linewidth]{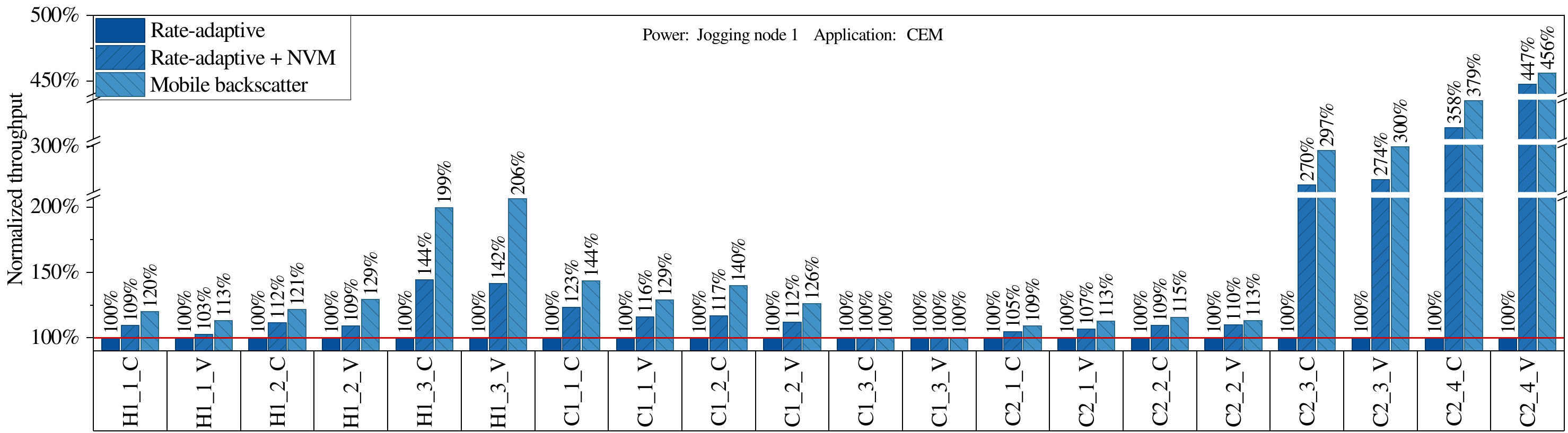}
\caption{Normalized throughput per signal trace for the \cem application under the jogging node 1 power trace.}
\label{fig:Improve_CEM_node1}
\vspace{-3mm}
\end{figure*} 

\fakepar{Throughput per trace}
\figref{fig:Improve_AEM}, \figref{fig:Improve_AR}, \figref{fig:Improve_CEM}, \figref{fig:Improve_AEM_node1}, \figref{fig:Improve_AR_node1}, and \figref{fig:Improve_CEM_node1} illustrate that the normalized throughput varies across different signal traces for all three applications under the jogging node 0 and jogging node 1 power traces.
Overall, these charts show that normalized throughput strongly depends on signal traces, while different applications exhibit similar patterns across power traces.
The mobile backscatter system achieves its highest throughput under the \textit{C2\_4\_V} signal trace.
As shown in~\figref{fig:C2_4_V}, this trace contains a long period, 63.4\%, of weak signal strength below the transmission threshold.
During these periods, the mobile backscatter system stores generated packets in NVM, and later exploits the RSSI trend to transmit them under improved channel conditions, leading to significant throughput gains.
In contrast, the NVM-based rate-adaptive baseline stores packets in NVM, but it does not exploit channel trends and energy status, resulting in lower throughput.
The rate-adaptive baseline simply discards packets when transmission is not possible, leading to the lowest throughput.

An exception is observed for the \ar application under the jogging node 1 power trace and the \textit{C2\_4\_V} signal trace, as shown in~\figref{fig:Improve_AR_node1}, where the mobile backscatter system performs worse than the NVM-based rate-adaptive baseline.
Because \ar has a lightweight computational workload and the jogging node 1 provides high input energy, the system generates a large number of packets during weak signal conditions.
However, \textit{C2\_4\_V} lacks sufficient transmission opportunities, as 63.4\% of the time the RSSI remains below the transmission threshold.
As a result, the mobile backscatter system retains packets in NVM to avoid inefficient transmissions.
In contrast, the NVM-based rate-adaptive baseline aggressively transmits packets whenever the minimal signal strength and energy requirements are met, regardless of channel quality.
While this yields higher short-term throughput, it is less energy-efficient.
Similar limitations exist under the \textit{H1\_3\_C} and \textit{H1\_3\_V} signal traces, where insufficient transmission opportunities limit the benefit of the mobile backscatter system.

Across all applications and power traces, the lowest throughput is observed under the \textit{C1\_3\_C} and \textit{C1\_3\_V} signal traces.
These traces are recorded when the tag moves around within a short and fixed distance (around 3 meters) from the carrier emitter, resulting in RSSI fluctuating between \mbox{-$20$\,dBm} and -30\,dBm, without a clear trend, as shown in \figref{fig:C1_3}.
Consequently, the system rarely encounters weak-channel periods that require deferring transmission, and the lack of a clear signal trend limits the effectiveness of the transmission control.
Thus, the mobile backscatter system performs similarly to the baselines.
The \ar application exhibits slightly lower throughput than the baselines, as its lightweight workload makes the overhead more noticeable.
Overall, in scenarios lacking clear signal trends, the mobile backscatter system offers comparable or slightly reduced performance compared to the baselines.

\begin{figure*}[tb]
\centering
\includegraphics[width=1\linewidth]{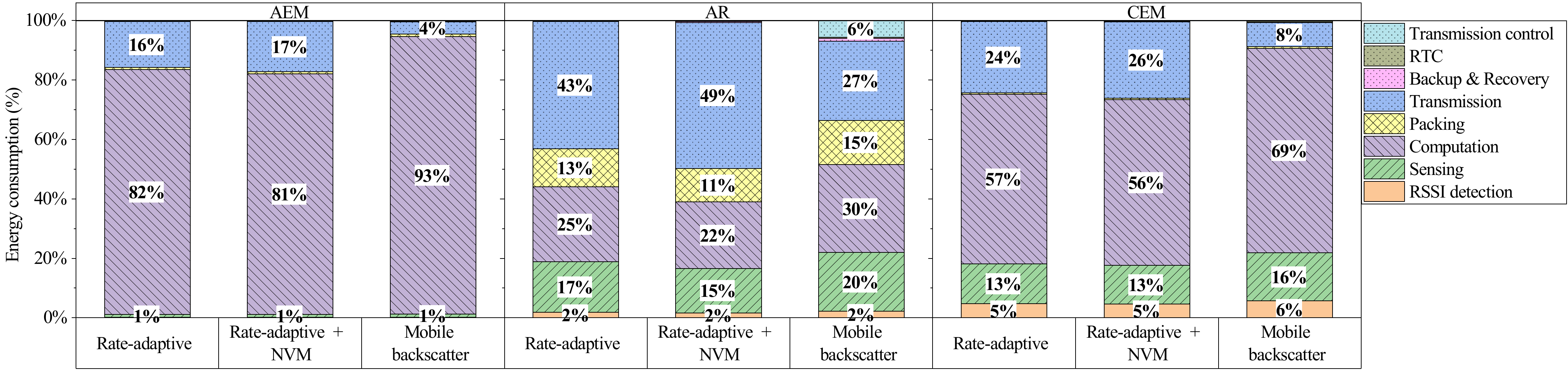}
\caption{Energy breakdown per application under the \textit{C1\_2\_C} signal trace and jogging node 0 power trace.
\capt{Transmission energy cost is significantly reduced in the mobile backscatter system.}}
\label{fig:energy}
\vspace{-3mm}
\end{figure*}

\fakepar{Throughput under different RSSIs}
\figref{fig:liftup} shows the number of transmitted packets across signal-strength levels for the \aem application under the \textit{C1\_2\_C} signal trace and jogging node 0 power trace.
The rate-adaptive baseline transmits all packets immediately based on the signal strength shown in the purple bar.
All three approaches exhibit similar immediate-transmission patterns, largely determined by the signal trace.
The NVM-based rate-adaptive baseline stores packets in NVM when the signal strength falls below the transmission threshold, but transmits them as soon as minimal energy and signal requirements are met.
Consequently, packets generated under weak signal conditions below -46\,dBm, highlighted in yellow, are primarily transmitted between -45\,dBm and -42\,dBm.

In contrast, the mobile backscatter system shifts these transmissions to significantly stronger signal regions, between -37.5\,dBm and -28\,dBm.
In addition, the transmission control stores packets even at usable RSSI for later transmission under favorable conditions.
These packets are highlighted in~\figref{fig:liftup} using distinct colors and patterns.
Finally, the mobile backscatter system transmits 619 packets, significantly outperforming the rate-adaptive baseline with 454 packets and the NVM-based rate-adaptive baseline with 555 packets.
This gain is achieved by deferring packets generated under poor channel conditions and transmitting them under stronger channel conditions, thereby reducing energy consumption per packet and improving throughput.

\begin{figure}[tb]
\centering
\includegraphics[width=1\linewidth]{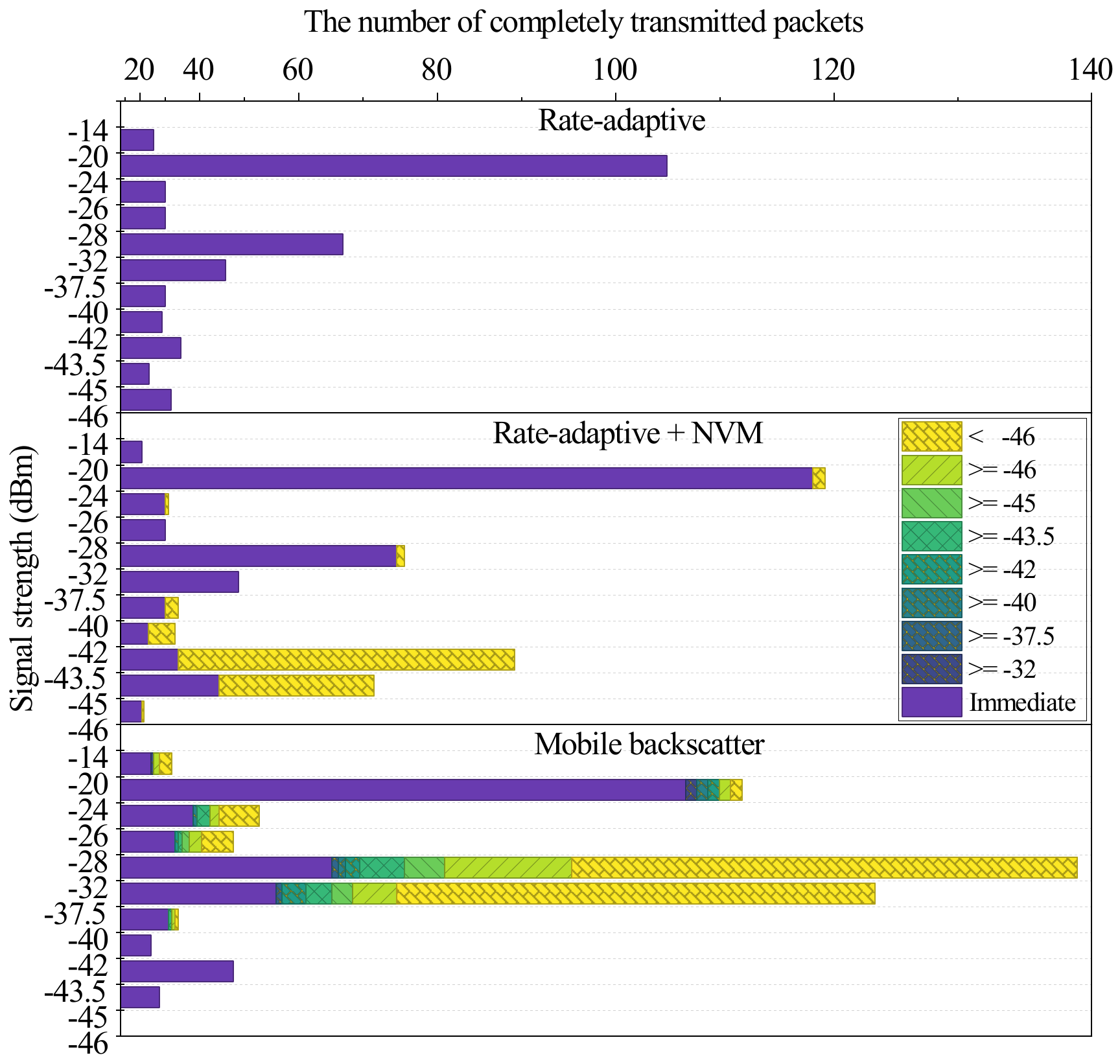}
\caption{Comparison of packet transmission behavior across RSSIs for the \aem application under the \textit{C1\_2\_C} signal trace and jogging node 0 power trace. 
\capt{Buffered packets are transmitted under more favorable channel conditions in the mobile backscatter system.}}
\label{fig:liftup}
\vspace{-3mm}
\end{figure}

\subsection{Energy}
\label{sec:energy}
We analyze task-level energy consumption to explain how the mobile backscatter system improves throughput compared to the baselines.
\figref{fig:energy} presents this breakdown across the three applications, using the \textit{C1\_2\_C} signal trace and jogging node 0 power trace as an illustrative example.

As discussed in~\secref{sec:applications}, \aem is a computation-intensive application.
\figref{fig:energy} shows that computation accounts for over 80\% of its total energy.
For the rate-adaptive baseline, only 16\% of the energy is consumed by transmissions.
By aggressively transmitting under favorable channel conditions, the mobile backscatter system reduces transmission energy to just 4\%, contributing to 12\% savings in transmission energy consumption.
This allows the system to allocate more energy to other tasks and generate more packets.
As a result, it achieves 1.36\,$\times$ the throughput of the rate-adaptive baseline.

\ar has a lighter computational and communication demand than the \aem.
Computation accounts for around 25\% of the total energy, while other tasks such as sensing and packet processing contribute another 30\%.
Because lighter workloads allow the system to generate more packets, energy demand shifts toward transmission, which dominates the energy consumption of both baselines, accounting for 43\% and 49\%, respectively.
The mobile backscatter system efficiently reduces this transmission energy to 27\%, achieving 1.36\,$\times$ the throughput of the rate-adaptive baseline, while reducing transmission energy consumption by 16\%.

\cem has a larger packet size but transmits less frequently because it compresses data until a block is full, as discussed in~\secref{sec:applications}.
This behavior results in a larger proportion of energy spent on RSSI monitoring.
By deferring transmissions during poor channel conditions, the mobile backscatter system reduces transmission energy to 8\%, compared to 24\% and 26\% for the baselines, respectively.
Consequently, it achieves 1.50$\times$ the throughput of the rate-adaptive baseline.

Across all evaluation results, the mobile backscatter system achieves up to 47.3\% reduction in transmission energy consumption compared with the rate-adaptive baseline.
While improving transmission energy efficiency, the system also introduces energy overhead from the transmission control, which must be quantified.
In \aem and \cem, this overhead is negligible, accounting for only 0.32\% and 0.23\% of the total energy consumption, respectively.
This is because the other tasks in the application dominate the energy budget.
In \ar, the lightweight workload and small packet size make this overhead more noticeable, consuming around 6\% of total energy. 
Nevertheless, the energy overhead remains limited, with a maximum of 7.3\% across all evaluated configurations.

\subsection{Packet Age}
\label{sec:age}
We also evaluate packet age, defined as the time between packet generation and transmission, to understand how the mobile backscatter system affects time-sensitive applications.

\begin{figure}[tb]
\centering
\includegraphics[width=0.85\linewidth]{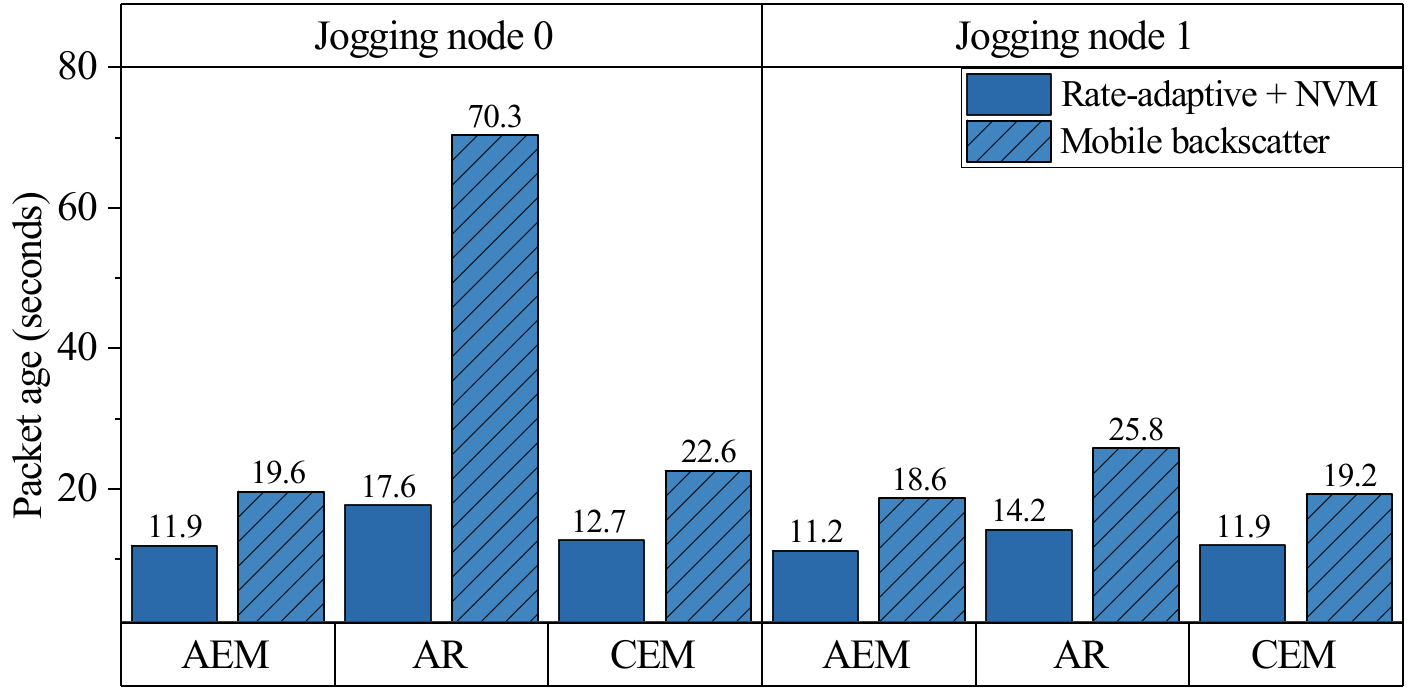}
\caption{Average packet age across three applications under all signal and power traces.}
\label{fig:Age}
\vspace{-3mm}
\end{figure}

\figref{fig:Age} illustrates the average packet age for the three applications across all signal and power traces.
The rate-adaptive baseline transmits packets immediately based on the instantaneous signal strength and is therefore not shown in this chart.
The NVM-based rate-adaptive baseline increases the average packet age to 13.25 seconds.
The mobile backscatter system results in a longer packet age.
It stores packets in NVM not only when transmission is not possible, but also proactively defers transmission based on the transmission control.
The system then selectively schedules stored packets for transmission under more favorable channel conditions.
As a result, packets may remain in NVM for longer periods, leading to an increased average packet age of 29.35 seconds.
The mobile backscatter system exhibits a significantly higher packet age in the \ar application under the jogging node 0 power trace. 
This is mainly due to the \textit{H1\_3\_C} and \textit{H1\_3\_V} signal traces, which result in average packet ages of 338 and 505 seconds, respectively, due to limited transmission opportunities.

\subsection{Micro-benchmarks}
\label{sec:microbench}
The throughput improvements of the mobile backscatter system depend on a few external dynamics. 
We investigate their impact in detail.

\begin{figure}[tb]
\centering
\includegraphics[width=0.9\linewidth]{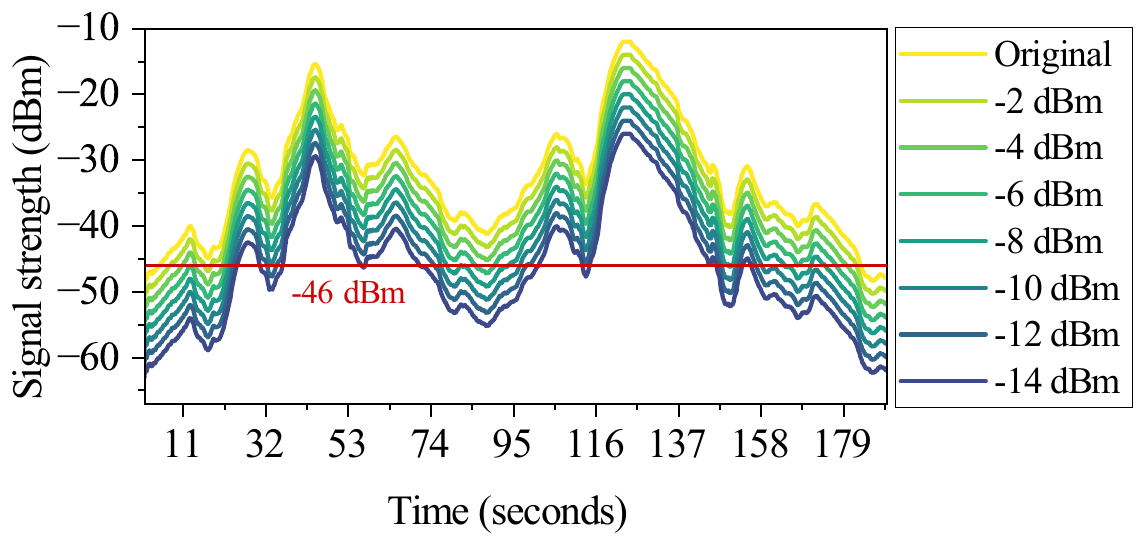}
\caption{Original and downward-shifted \textit{C2\_2\_C} signal traces.}
\label{fig:shifts}
\vspace{-3mm}
\end{figure} 

\begin{figure}[tb]
\centering
\includegraphics[width=1\linewidth]{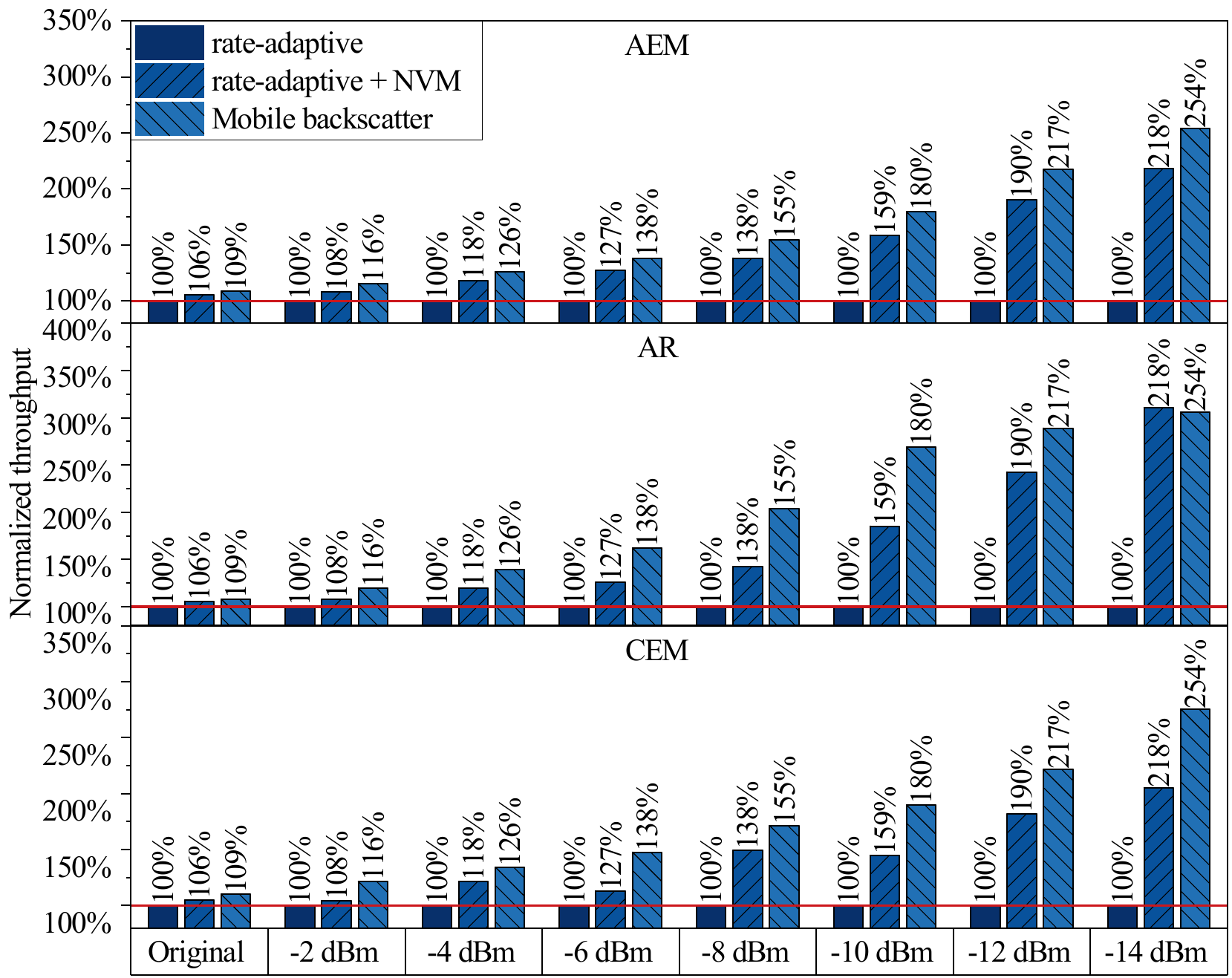}
\caption{Normalized throughput under \textit{C2\_2\_C} signal traces with different RSSI offsets. 
\capt{The mobile backscatter system increases throughput as RSSI offsets become larger.}}
\label{fig:shifts_throughput}
\vspace{-3mm}
\end{figure} 

\fakepar{Signal traces}
To further isolate the impact of signal variations, we conduct a controlled study by shifting the \textit{C2\_2\_C} signal trace downward in steps of 2\,dBm as shown in~\figref{fig:shifts}.
\figref{fig:shifts_throughput} presents the corresponding throughput results under the jogging node 0 power trace.

As the downward offset increases, the mobile backscatter system achieves higher throughput compared to both baselines. 
In the original trace, signal strength falls below the transmission threshold only 7\% of the time, providing limited opportunities to store packets in NVM.
As the signal strength is reduced by increasing the offset, weak signal periods become longer. 
Both the NVM-based rate-adaptive and mobile backscatter systems store more packets in NVM. 
However, the mobile backscatter system achieves higher throughput by leveraging signal trends to defer these buffered packet transmissions under weak channel conditions.

An exception occurs in the \ar application at a -14\,dBm offset, where our design achieves lower throughput than the NVM-based rate-adaptive baseline. 
This is because many packets are generated during periods of weak signal, leading to a large number of packets stored in NVM.
Simultaneously, the -14\,dBm offset reduces the proportion of favorable channel conditions, thereby decreasing transmission opportunities.
Nevertheless, if favorable channel conditions become available later, the mobile backscatter system can still clear the stored packets and outperform the NVM-based rate-adaptive baseline.

\begin{figure}[tb]
\centering
\includegraphics[width=0.8\linewidth]{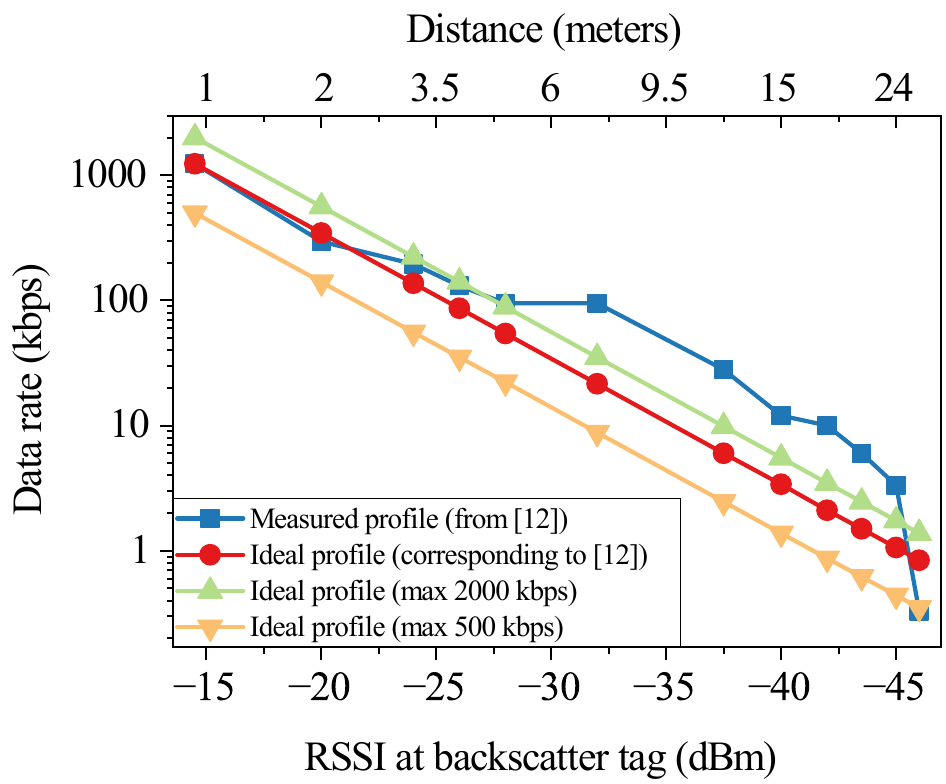}
\caption{Different rate-RSSI profiles.}
\label{fig:Curves}
\vspace{-3mm}
\end{figure} 

\begin{figure}[tb]
\centering
\includegraphics[width=1\linewidth]{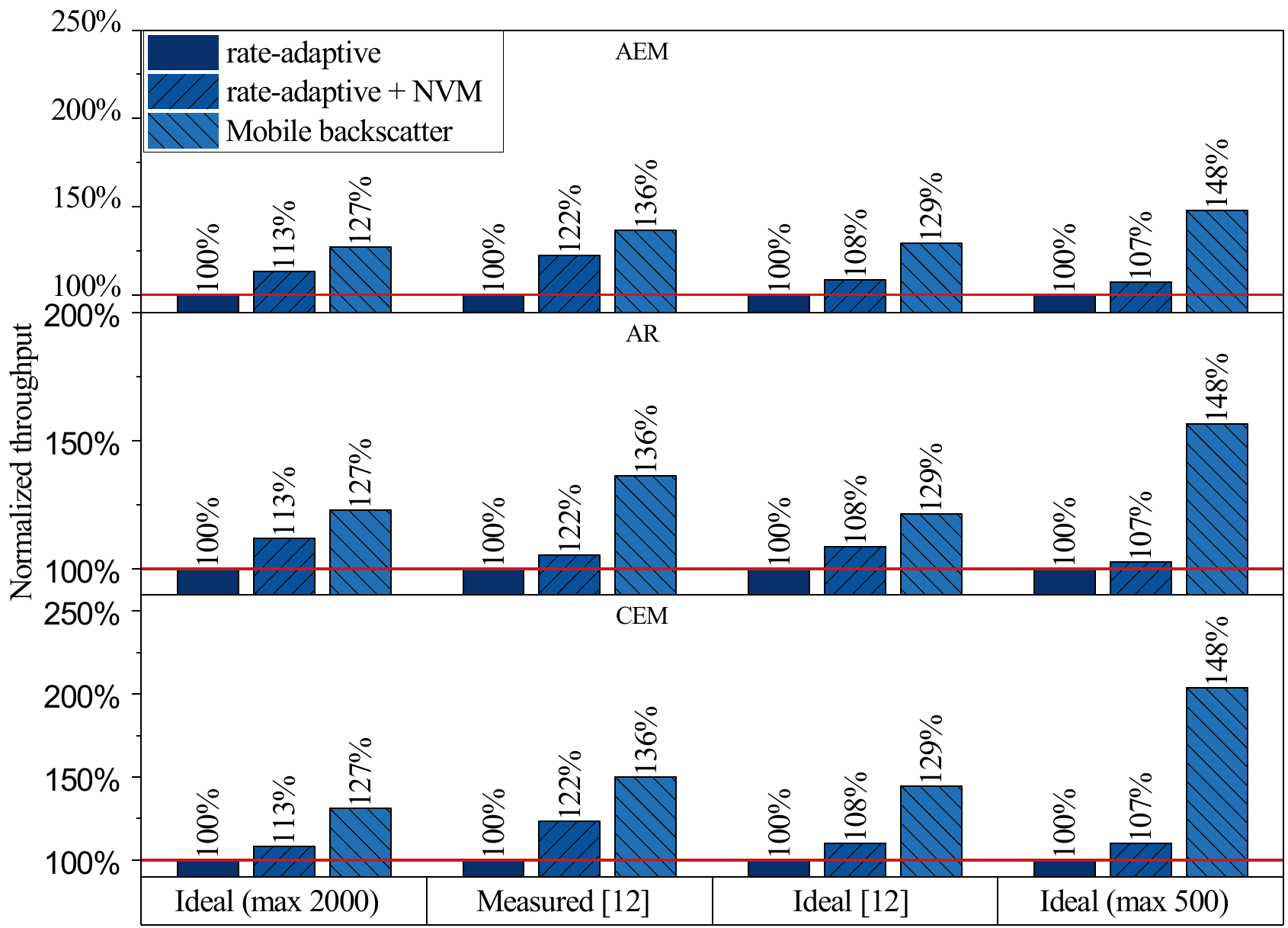}
\caption{Normalized throughput under \textit{C1\_2\_C} signal trace and jogging node 0 power trace across rate-RSSI profiles. 
\capt{The mobile backscatter system achieves higher throughput with lower maximum data rates across different rate–RSSI profiles.}}
\label{fig:Curves_throughput}
\vspace{-5mm}
\end{figure} 

\fakepar{Impact of the rate–RSSI profile}
The rate–RSSI profile used in our system is from prior work~\cite{huang2021freeback} and replicated in our experimental setup. 
To understand its impact on performance, we evaluate four rate–RSSI profiles, as shown in~\figref{fig:Curves}.
The blue curve is the replicated profile from the prior work~\cite{huang2021freeback}. 
The red curve corresponds to an ideal scenario without multipath effects, such as reflections from walls or surrounding objects.
In addition, we construct two idealized profiles with maximum data rates of 500\,kbps and 2000\,kbps, respectively.

\figref{fig:Curves_throughput} presents the throughput results.
The mobile backscatter system achieves higher throughput when the maximum data rate is lower. 
This is because lower data rates under weak signal conditions result in larger energy consumption, making it more beneficial to exploit better channel conditions.
Furthermore, the mobile backscatter system achieves comparable throughput under the idealized profile compared to the measured profile replicated from prior work~\cite{huang2021freeback}.
In contrast, the NVM-based rate-adaptive baseline achieves lower throughput under the idealized profile.
This is because the idealized profile exhibits lower average data rates, leading to higher energy consumption, which reduces throughput. 
Despite this, the mobile backscatter system still achieves consistent gains by aligning transmissions with favorable channel conditions.

\subsection{Adaptability Under Dynamic Conditions}

With prior knowledge of the environment, transmission policies can be tuned beforehand to specific channel and energy conditions. 
For example, knowing carrier emitter placement, mobility patterns, and energy intakes would allow offline optimization of transmission decisions. 
In this setting, even simple transmission rules may achieve high throughput after tuning.
In practice, however, these aspects are fundamentally unpredictable, making offline optimization impractical.

In contrast, our approach adapts transmission decisions online based on observed RSSI trends and energy availability, exploiting transmission opportunities under changing conditions without relying on environment-specific tuning.

\section{Conclusion}
\label{sec:conclusion}

We presented a mobile backscatter communication system for battery-less mobile IoT devices.
Our design introduces a lightweight transmission control that regulates packet transmissions based on real-time channel and energy variations.
We persist critical state and packets in NVM, maintaining progress across energy failures and poor channel conditions where transmissions are inefficient or impossible.
We implement this design on a real hardware prototype based on the MSP430FR5969 MCU and evaluate the system using three representative IoT applications and real-world power and signal traces. 
Experimental results show that we achieve up to 5.16× the throughput of the rate-adaptive baseline, enabled by up to 47.3\% reduction in transmission energy consumption, while introducing only 0.23\% -- 7.3\% additional energy overhead.

\fakepar{Acknowledgements} This work is supported by the Swedish Foundation for Strategic Research (SSF) and by the National Recovery and Resilience Plan (NRRP), Mission 4 Component 2 Investment 1.3—Call for Tender No. 1561 of 11.10.2022 of the Italian Ministero dell’Università e della Ricerca (MUR); funded by the European Union—NextGenerationEU.

\bibliographystyle{IEEEtran}
\bibliography{references}

\end{document}